# Slow Light Enhanced Photon Echoes

Joonseong Hahn and Byoung S. Ham[*]

*Center for Photon Information Processing, and the Graduate School of Information & Telecommuications, Inha University, 253 Yonghyun-dong, Nam-gu, Incheon 402-751, S. Korea*

[*]*bham@inha.ac.kr*

**Abstract:** We report a slow light-enhanced photon echo method, whose retrieval efficiency is two orders of magnitude higher than that of conventional photon echoes. The enhanced photon echo efficiency is due to lengthened interaction time given by ultraslow group velocity.

Since the first observation of reversible macroscopic atomic coherence[1], photon echoes have been intensively studied for ultrahigh speed all-optical memories[2-4] as well as multimode quantum memories[5-8]. A major problem of photon echoes is a contradictory effect between data absorption and echo emission resulting in low retrieval efficiency. Modified photon echo techniques can enhance the low retrieval efficiency, but at the cost of Raman scattering, complicated design, or a reduced bandwidth[5-8]. The ultraslow light[9] using electromagnetically induced transparency (EIT) (ref. 10) has attracted much attention to quantum memories by mapping a flying quantum bit of light into an atomic ensemble: however an inherently low data rate limits potential and practical applications[11-14].

Photon echoes[1] have been intensively studied for ultrafast and ultradense optical memories since the 1980s (refs. 2-4). Even though several techniques have been developed to increase the storage density and data rate, extremely low retrieval efficiency has been a main drawback in photon echo implementation. In conventional photon echoes, retrieval efficiency is less than 1% in most rare-earth doped solid media. Recently, modified photon echo methods have improved photon echo efficiency[5-8]. These methods include applying frequency comb-based phase grating[5], using spin coherence interactions in backward three-pulse photon echo methods[6,7], and adapting an electric DC field to replace a π optical pulse[8].

In this Letter, we report slow light-enhanced photon echoes whose efficiency is two orders of magnitude higher than that of conventional photon echoes. The configuration of the present method is as simple as the conventional photon echo methods, except for its use of dummy light for slow light preparation. The dummy light, whose frequency is the same as the data pulse, plays a key role in the enhanced photon echo efficiency by burning a spectral hole in an inhomogeneously broadened optical medium, so that following optical data pulses experience ultraslow group velocity due to modified spectral redistribution. The enhanced photon echo efficiency results from the lengthened interaction time between photons and the optical medium under the ultraslow group velocity. We analyze the enhanced photon echo efficiency using the slow light phenomenon and discuss potential applications of quantum memories.

In conventional photon echoes, optical data pulses excite inhomogeneously broadened atoms from the ground state to the excited state, where the information of the optical data pulses is imprinted as a superposition of spectral modulations (or phase gratings) of the ground state atoms as well as the excited atoms. This multimode storage capability in time domain, whose maximum data rate is the same as the inhomogeneous width, is the most important benefit of photon echoes to quantum memory applications. Before the homogeneous decay time of the excited atoms, a π pulse, the so-called rephasing pulse, is applied to flip the phase information (uv plane in a Bloch vector model) of each of the excited atoms. Then all atoms start to rephase by themselves in a reversed time manner. Even though the highly absorptive medium should be advantageous to the photon echo excitation, an exponential drop of absorption in **k** space according to Beer's law causes exponential reabsorption of the generated photon echoes[15]. In other words, the stronger the photon echo excitation, the stronger the photon echo reabsorption. This outcome represents a fundamental dilemma in conventional photon echoes.

Over the last decade ultraslow light has been an important research topic in nonlinear quantum optics owing to enhanced wave mixing processes due to lengthening interaction[9]. Thus, ultraslow light has sought its potential applications in quantum information processing such as entanglement generation[16] and quantum memories[11-14]. Although quantum memory based on ultraslow light has been intensively studied for a promising protocol, the inherently low single mode data rate has presented a fundamental drawback to practical quantum information processing[17,18]. The interaction time in ultraslow light, however, is a controllable parameter, so that low power nonlinear optics[19] has evolved for potential applications of both classical and quantum information. Recent observation of a dummy light-triggered ultraslow light offers a great benefit to the ultraslow light applications owing to lengthened interaction time as well as a simple configuration using the same frequency[20]. As noted above, the main dilemma of photon echoes − the contradictory requirement of an optically dense medium − can be solved if a dummy light-triggered ultraslow light is applied, because longer interaction time can compensate a weaker absorption rate even in an optically light medium. This dilemma is the motivation of the present work and will be discussed below.

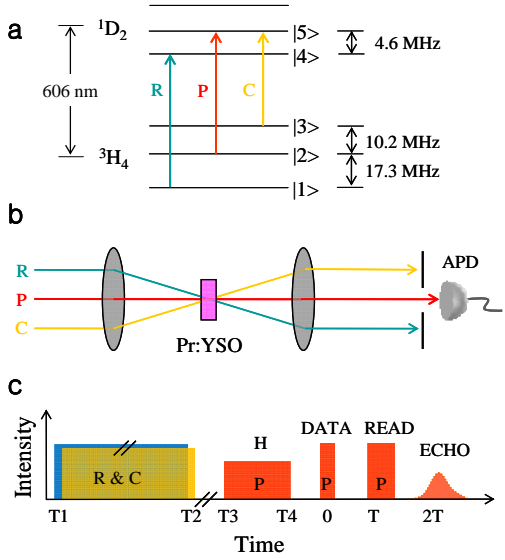

Figure 1. Schematic of ultraslow light enhanced photon echo. (a) partial energy level diagram of Pr:YSO interacting with laser light. (b) Pulse propagation. APD: avalanche photo diode, R and C: repump light. P: dummy, data, and rephrasing pulse. (c) Pulse sequence. T1=−6.6 ms; T2=−1.6 ms; T3=−1.1 ms; T4=−0.5 ms; T=7 μs.

Figure 1 shows a schematic of the present ultraslow light enhanced photon echoes. Figure 1a shows a partial energy level diagram of 0.05 at. % $Pr^{3+}$ doped $Y_2SiO_5$ (Pr:YSO) interacting with laser light R, C, and P. The light pulses R and C play a repumping role to transfer atoms from states |1> and |3> to state |2> to increase atom density for P transition. Thus Fig. 1a turns out a two-level system for the transition of P. The light beams R and C are set not to be collinear with P to prevent any unwanted noise from the echo detection. As shown in Fig. 1c, the light P is decomposed into three pulses: H, DATA, and READ. The light H acts as dummy light for the preparation of ultraslow light by burning a spectral hole (see ref. 20). Thus, the subsequent DATA/READ pulses experience ultraslow group velocity. The group velocity of the DATA/READ pulse can easily be controlled by the power R, C, and/or H as demonstrated in Ref. 20. Under the ultraslow light, the DATA pulse absorption increases strongly (shown in Fig. 2). However, the echo emission experiences less reabsorption due to the atom depletion by H-induced spectral hole burning. The length $l$ of the medium is 5 mm, and the measured optical depth d (d=$\alpha l$, where $\alpha$ is absorption coefficient) with the repump R and C, but without the dummy light H, is about 20.

The light pulses in Fig. 1 are from a ring-dye laser (Tecknoscan) via acoustooptic modulators (Isomet) driven by digital delay generators (Stanford DG 535) and radiofrequency synthesizer (PTS 250).The angle among R, C, and P is ~12 mrad and overlapped by 80% through the sample. The optical signals detected by an avalanche photon diode are recorded in the oscilloscope by averaging 30 samples. The repetition rate of the light pulse sequence is 20 Hz. The temperature of the sample is kept at ~5 K. The power of R, C, H, P is 42, 45, 25, and 7 mW, respectively. The pulse length of R, C, and H is 5 ms, 5 ms, and 600 μs. The absorption rate for P transition is several times higher than the others. The light is vertically polarized and the crystal axis of the medium, Pr:YSO is along the propagation direction **k**. The temperature dependent homogeneous width is measured by scanning frequency and measuring the DATA pulse only.

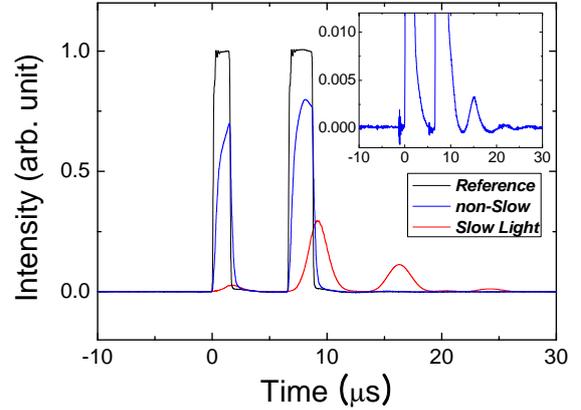

Figure 2. Photon echoes using slow light (red curve) and nonslow light (blue curve). Black curve is for the consecutive two input pulses of data and rephsing as a reference. Pulse length of DATA and READ is 1.5 μs and 2.3 μs, respectively. The power of P is 11 mW.

Figure 2 shows experimental data of the ultraslow light-enhanced photon echoes. The pulse length of the DATA and READ is 1.5 μs and 2.3 μs,

respectively. The separation between DATA and READ is 5 μs. The black line represents a reference of the input pulse intensity (DATA and READ) before entering the medium, Pr:YSO. The red and blue curves are the intensity of the transmitted light with and without the dummy light H, respectively. The inset of the expanded blue curve shows a typical conventional photon echo signal without applying the H field. As seen in the inset, the photon echo efficiency (intensity ratio of the echo to the reference DATA) is 0.3%, while the slow light enhanced photon echo efficiency (red curve) is measured at 12%. The DATA pulse absorption in the slow light regime is much higher than that of conventional photon echoes due to lengthened interaction time (see Fig. 3). From this we calculate that the enhancement factor with the ultraslow light effect is 40.

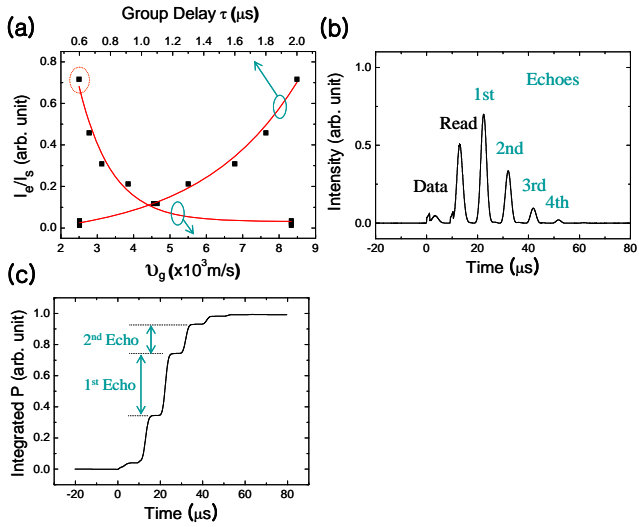

Figure 3. Slow light enhanced photon echoes. (a) Echo intensity $I_e$ versus group delay (velocity). (b) Slow light enhanced photon echo for the 2 μs group delay in (a). (c) Integrated intensity of (b). Data pulse intensity represents 1.0. Power of P is 11 mW.

Figure 3a shows echo intensity versus group delay (group velocity), where the echo intensity increases exponentially as the group delay increases. We adjusted the group delay by controlling the intensity $I_R$ of the repump R, where $I_R$ determines atom number N in state $|2\rangle$: $N \propto Sin(\sqrt{I_R}T)$, and T is the pulse duration of $I_R$ (see Fig. 2 of ref. 20). Figure 3a confirms that the exponential increase of the echo intensity (as a function of the group delay) results from increased interaction time of the data pulse. Figure 3b represents detailed data of Fig. 3a for the slowest group velocity (see the red-dotted circles in Fig. 3a), where multiple echoes are emitted owing to the strong atomic coherence process. All pulse intensities in Fig. 3b are integrated to denote echo efficiency, where the total transmission of Fig. 3b is 79% of the input DATA pulse. The first echo efficiency is as high as 31%, which is more than 100 times that of the conventional photon echo shown in Fig. 2. The cumulative echo efficiency reaches 51% of the DATA pulse. The lost 21% is due to absorption of high frequency components and scatterings including Raman processes.

In Fig. 4a, the slow light enhanced photon echo intensity obtained in Fig. 2 is plotted as a function of delay time T of the READ pulse. From the best-fit exponential curve the decay time τ is 25 μs: $I_{echo} = I_0 \exp(-2t/\tau)$. According to the photon echo theory, the two- (three-) pulse echo measures optical phase (population) decay time. Surprisingly the phase decay time obtained in Fig. 4a is too short compared with that in ref. 21 ($T_2$=111 μs at 2K). This shortened decay time may be due to a temporary heat-up by the H pulse, where the Pr:YSO sample is barely contacted with the cold finger of the cryostat. In our record, the measured $T_2$−dependent optical homogeneous broadening varies from kHz at ~2 K to 10 MHz at ~10 K. This means that the homogeneous decay time can be severely shortened, even with a one degree change in temperature.

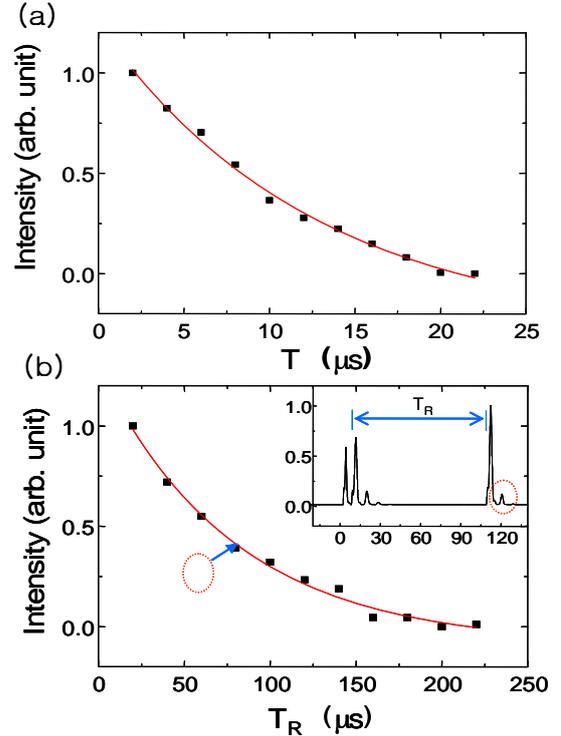

Figure 4. Photon echo intensity versus delay time of READ pulse. (a) two-pulse photon echo. Power of P is 11 mW. (b) three-pulse photon echo. Inset: experimental data.

In Fig. 4b the READ pulse used in Fig. 4a is divided into two identical pulses (WRITE and READ) for the test of a typical three-pulse photon echo experiment to measure optical population relaxation time $T_1$. The echo intensity is measured as a function of $T_R$ (separation between two $\pi/2$ identical rephasing pulses) for a fixed T (T=10 μs). From the best-fit exponential curve, we calculate that the measured decay time (population decay time) is 158 μs, which is very close to the temperature independent optical population decay time $T_1$, ($T_1$ = 164 μs in ref. 21). Moreover the present slow light enhanced photon echo also fulfills the rephasing mechanism (see Supplementary Fig. S1). The inset shows a typical three-pulse photon echo among the given data (red dotted circle), where the last peak represents the three-pulse photon echo signal. Thus, Fig. 4 supports the generality of photon echo theory in the ultraslow light regime.

In conclusion we experimentally demonstrated and analyzed slow light-enhanced photon echoes in a rare earth $Pr^{3+}$ doped $Y_2SiO_5$. The observed echo efficiency is two orders of magnitude higher than that of conventional photon echoes. This observation holds potential for quantum memory applications using the same technique of photon echoes by simply adding a dummy light, whose frequency is the same as the data pulse. Compared with other modified photon echo methods, the present technique is simpler and keeps the same benefit of photon echoes: multi-dimensional, ultrafast all-optical quantum memories.

**Acknowledgement**

This work was supported by the Creative Research Initiative Program (Center for Photon Information Processing) of the Korean government (MEST) via National Research Foundation of Korea.